\documentstyle[12pt]{article}

\begin{document}

\def\m{{\bf m}}
\def\n{{\bf n}}
\def\r{{\bf r}}
\def\v{{\bf v}}
\def\x{{\bf x}}
\def\F{{\bf F}}
\def\R{{\bf R}}
\def\Z{{\bf Z}}
\def\del{\partial}
\def\Lap{\bigtriangleup}
\def\Div{{\rm div}\ }
\def\rot{{\rm rot}\ }
\def\curl{{\rm curl}\ }
\def\grad{{\rm grad}\ }
\def\^{\wedge}
\def\real{{\rm re}}
\def\image{{\rm im}}
\def\goinf{\rightarrow\infty}
\def\goes{\rightarrow}
\def\bm{\boldmath}
\def\eV{{\rm eV}}
\def\keV{{\rm keV}}
\def\MeV{{\rm MeV}}
\def\GeV{{\rm GeV}}
\def\TeV{{\rm TeV}}
\def\-{{-1}}
\def\inv{^{-1}}
\def\wa{\!\!\!\!&=&\!\!\!\!}
\def\wb{\!\!\!\!&\equiv &\!\!\!\!}

\def\eqn#1#2{\if#2 \begin{equation} #1 \end{equation} \else\begin{equation}
 #1 \label{#2} \end{equation} \fi}
\def\eqna#1{\begin{eqnarray} #1 \end{eqnarray}}
\def\reff#1{(\ref{#1})}
\def\vb#1{{\partial \over \partial #1}} 
\def\Del#1#2{{\partial #1 \over \partial #2}}
\def\Dell#1#2{{\partial^2 #1 \over \partial {#2}^2}}
\def\Dif#1#2{{d #1 \over d #2}}
\def\Lie#1{ {\cal L}_#1 }
\def\abs#1{\left | #1 \right |}
\def\rcp#1{{1\over #1}}
\def\paren#1{\left( #1 \right)}
\def\brace#1{\left\{ #1 \right\}}
\def\bra#1{\left[ #1 \right]}
\def\angl#1{\left\langle #1 \right\rangle}
\def\lvector#1#2#3#4{\paren{\begin{array}{c} #1 \\ #2 \\ #3 \\ #4
\end{array}}}
\def\vector#1#2#3{\paren{\begin{array}{c} #1 \\ #2 \\ #3 \end{array}}}
\def\svector#1#2{\paren{\begin{array}{c} #1 \\ #2 \end{array}}}
\def\matrix#1#2#3#4#5#6#7#8#9{
	\left( \begin{array}{ccc}
			#1 & #2 & #3 \\	#4 & #5 & #6 \\ #7 & #8 & #9
	\end{array}	\right) }
\def\smatrix#1#2#3#4{
	\left( \begin{array}{cc} #1 & #2 \\	#3 & #4 \end{array}
\right) }
\def\manbo#1{}

\def\GL#1{{\rm GL}(#1)}
\def\SL#1{{\rm SL}(#1)}
\def\PSL#1{{\rm PSL}(#1)}
\def\O#1{{\rm O}(#1)}
\def\SO#1{{\rm SO}(#1)}
\def\IO#1{{\rm IO}(#1)}
\def\ISO#1{{\rm ISO}(#1)}
\def\U#1{{\rm U}(#1)}
\def\SU#1{{\rm SU}(#1)}

\def\d{{\rm d}}
\def\diag#1{{\rm diag}(#1)}
\def\uh#1#2{h^{#1#2}}
\def\dh#1#2{h_{#1#2}}
\def\ug#1#2{g^{#1#2}}
\def\dg#1#2{g_{#1#2}}
\def\E#1#2{E^{#1}{}_{#2}}
\def\psibari#1#2{\bar{\psi}^{{#1}'}{}_#2}
\def\psii#1#2{{\psi}^{#1 }{}_#2}

\def\HK{1}
\def\RS{2}
\def\GrPRL{3}
\def\GrPL{4}
\def\DHO{5}
\def\DE{6}
\def\DEpre{7}
\def\HH{8}
\def\HP{9}
\def\BGPP{10}
\def\GP{11}
\def\DErev{12}
\def\OSB{13}
\def\KTH{14}

\def\refmark#1{{[#1]}}

\def\abstract#1{\begin{center}{\large ABSTRACT}\end{center} \par #1}
\def\title#1{\begin{center}{\large #1}\end{center}}
\def\author#1{\begin{center}{\sc #1}\end{center}}
\def\address#1{\begin{center}{\it #1}\end{center}}
\def\pubnum{221/COSMO-30}

\begin{titlepage}
\hfill
\parbox{6cm}{{TIT/HEP-\pubnum} \par May 1993 }
\par
\vspace{7mm}
\title{Supersymmetric Bianchi class A models}
\vskip 1cm
\author{ Masako ASANO \footnote{E-mail address: maa@phys.titech.ac.jp},
Masayuki TANIMOTO \footnote{E-mail address: prince@phys.titech.ac.jp}
and Noriaki YOSHINO
}
\address{Department of Physics, Tokyo Institute of \\ Technology, Oh-Okayama
Meguroku, Tokyo 152, Japan}
\vskip 1 cm
\abstract{
The canonical theory of $N=1$ supergravity is applied to Bianchi class A
 spatially homogeneous cosmologies.
The full set of quantum constraints are then solved with the possible
ordering ambiguity taken into account by introducing a free parameter.
The wave functions are explicitly given for all the Bianchi class A models
in a unified way.
Some comments are made on the Bianchi type IX cases.
}

\vfill
\end{titlepage}

\addtocounter{page}{1}

\section{Introduction}

Recently, supersymmetry has been attracting many people in particle
physics (see, e.g., ref. [\HK]).
It seems worth studying supersymmetric quantum cosmology by expecting the
existence of supersymmetry in the early universe.

Mini-superspace models, i.e. spatially homogeneous relativistic models,
have been helpful to understand quantum cosmology.
Among them, those of Bianchi class A [\RS] (its definition explained in
section 2) are of particular importance when one investigates them in canonical
 form, because for Bianchi class `B' models, the rest of the Bianchi types,
there is an inconvenience that the equations of motion reduced from the
 full Einstein equations differ from the ones derived from the reduced
 action.
Since the origin of this disaccord is purely geometrical we expect the
 same problem will happen also in supergravity.
So we restrict our attention to the supersymmetric extension of the class
 A models.
Special cases of such models have been investigated by some authors.
Graham studied Bianchi type IX in refs. [\GrPRL,\GrPL], where the full quantum
 constraints are not explicitly solved.
D'Eath, Hawking and Obreg\'{o}n [\DHO]
\footnote{Our result presented below disagrees with [\DHO].}
solved the Bianchi type I case, but the Bianchi type I supersymmetric model
 is so simple that the wave function does not have any interesting structure.
\footnote{
After the first submission of this paper
it came to our attention that related work
 [\DErev,\OSB] by some other authors had appeared.
They treat the Bianchi type IX cases and the type II cases,
respectively.
Here, we emphasize it is worth studying all class A models even when
considering compactification [\KTH], contrary to the reason explained in
ref. [\DErev] for which the author did not study Bianchi models except types I
and IX.
}

We shall first restrict the canonical theory of supergravity proposed by
 D'Eath [\DE] to spatially homogeneous cases, in particular, Bianchi class A
 models,
and then solve the quantum constraints.
Then we will obtain the wave function
\eqn{\Psi =const. h^{\frac s{2}}\exp \paren{\frac 1 2 \gamma m^{ab}h_{ab}}+
const. h^{\frac {2-s}{2}}\exp \paren{-\frac 1 2 \gamma m^{ab}h_{ab}}(\psii
 Aa \psi_{Aa})^3,}{danii-1}
where $\dh ab$ is a spatial metric with respect to an invariant basis, $h$ its
 determinant and $m^{ab}$ the structure constants of a Bianchi class A group.
Here, $\gamma$ is a real constant and
$\psii Aa$ a spin-$3\over2$ field.
The parameter $s$ specifies the ambiguity of the operator ordering (see
eqs. (\ref{asano-24}),(\ref{asano-25})).
This is a direct generalization of the solutions found in the literature
 [\GrPRL,\GrPL,\DHO].

\section{Spatially homogeneous Bianchi class A models}

We write the four-metric as
\eqn{\d s^2=(N^iN_i-N^2)\d t^2+2N_i\d t\d x^i+\dh ij\d x^i\d x^j,}{1-1}
where $N$ is the lapse function, $N^i$ the shift vector, and
$\dh ij$ the spatial metric.
In spatially homogeneous cases, $\dh ij$ is expanded by an invariant basis
 of a Bianchi group,
\eqn{\dh ij=\dh ab(t)\E ai\E bj,}{1-2}
where indices $a,b,\dots$
as well as the world indices $i,j,\dots$ run from 1 to 3.
The invariant basis $\E ai$ obeys the following Maurer-Cartan relation:
\eqn{\del_{[i}\E a{j]}=C^a{}_{bc}\E bi\E cj,}{1-3}
where $C^a{}_{bc}$ are the structure constants of the Bianchi group.
$C^a{}_{bc}$ is antisymmetric in $b$ and $c$ and satisfies the Jacobi identity
$C^a{}_{b[c}C^b{}_{de]}=0$.
We can decompose $C^a{}_{bc}$ as
$C^a{}_{bc}=m^{ad}\epsilon_{dbc}+\delta^a_{[b}a_{c]}$,
where $m^{ad}$ is symmetric and $a_c\equiv C^a{}_{ac}$.
The Bianchi class A models are characterized by $a_c=0$, so that we have
\eqn{C^a{}_{bc}=m^{ad}\epsilon_{dbc}.}{1-5}

In supergravity, the basic variables are taken to be the spinor version of
 the tetrad components $e^{AA'}{}_\mu$ and the spinor valued forms
 $\psi^A{}_\mu,\bar\psi^{A'}{}_\mu$ (for spinor conventions, see ref. [\DE]).
$e^{AA'}{}_\mu$ are Grassmann even
and $\psi^A{}_\mu$ and $\bar\psi^{A'}{}_\mu$ are Grassmann odd.
In accordance with \reff{1-1}, we take
\eqn{e^{AA'}{}_0=Nn^{AA'}+N^ie^{AA'}{}_i.}{1-6}
Here, $n_{AA'}$ is the spinor version of the unit vector normal to the
 surface $t=$const., and it is defined by
\eqn{n_{AA'}e^{AA'}{}_i=0,\quad n_{AA'}n^{AA'}=1.}{1-7}
 From homogeneity the variables $N$, $N_i$, and $e^{AA'}{}_i$ are
 restricted to
\eqn{N=N(t),\quad N_i =N_a(t)E^a{}_i,\quad e^{AA^{\prime}}{}_i=
\Omega ^{AA'}{}_a(t)E^a{}_i.}{1-8}
The metric-like matrix $h_{ab}$ defined in \reff{1-2} is associated with
 $\Omega^{AA'}{}_a$ through
\eqn{h_{ab} = -\Omega_{AA'a}\Omega^{AA'}{}_b.}{1-9}
We will use $h_{ab}$ and its inverse $h^{ab}$ to raise and lower the
 indices $a,b,c,...$ .
Further, the spin- $\frac 3 2$ field $\psii A{\mu}$ will take the following
 spatially homogeneous form:
\eqn{\psii A0 =\psii A0 (t),\quad \psii Ai=\psii Aa(t) E^a{}_i.}{1-10}
Similar restrictions are to be imposed on $\psibari A{\mu}$.

\def\psibari#1#2{\bar{\psi}^{{#1}'}{}_#2}
\def\psii#1#2{{\psi}^{#1 }{}_#2}

The full theory of supergravity reduces to a theory with a finite number
 of degrees of freedom and its Lagrangian is given by the integration of
 the Lagrangian density ${\cal L}(t,x)$ over the spatial coordinates,
\eqn{L(t)=\int \d^3\! x {\cal L}(t,x),}{ }
with $N$, $N_a$, $\Omega ^{AA'}{}_a$, $\psii Aa$, $\psii A0$, $\psibari Aa$,
 and $\psibari A0$ being the variables of the reduced theory.
The reduced action $S \equiv \int \d t L(t)$ remains invariant under the
 three homogeneity preserving local transformations  -  Lorentz, coordinate,
and supersymmetry transformations.
Thus, it is possible to get the reduced Hamiltonian $H(t)$ by integrating
 the Hamiltonian density.
This results in a form formally equivalent to the original one:
\eqn{H(t)=NH_{\bot}+N^aH_a+\psii A0 S_A +\bar S_{A'}\psibari {A}
 0 -\omega _{AB0}J^{AB}-\bar{\omega}_{A'B'0}\bar{J}^{A'B'}.}{ }
Here, $N$, $N^a$, $\psii A0$, $\psibari A0$, $\omega_{AB0}$, and
 $\bar{\omega}_{A'B'0}$ act as Lagrange multipliers which enforce the
 constraints $H_{\bot}=0$, $H_a=0$, $S_A=0$, $\bar S_{A'}=0$, $J^{AB}=0$, and
 $\bar {J}^{A'B'}=0$.

Using the same procedure as shown in ref. [\DE], we can perform canonical
quantization of this system.
A quantum state is described by a wave function $\Psi$, which we choose
 as an eigenstate of $\psii Aa$ and $\Omega^{AA'}{}_a$, i.e.,
 $\Psi = \Psi (\Omega ^{AA'}{}_a, \psii Aa)$.
Corresponding to this choice, $\psibari Aa$ is given by
\eqn{\psibari Aa=-i\hbar D^{AA'}{}_{ba}\frac {\partial}{\partial \psii Ab}
\quad ,}{ }
and the momentum conjugate to $\Omega^{AA'}{}_a$, $p_{AA'}{}^a$, by
\eqn{p_{AA'}{}^a=-i\hbar \frac {\partial}{\partial \Omega ^{AA'}{}_a}+
\frac 1 2 i\hbar \sigma \epsilon ^{abc} \psii Ab D^{BA'}{}_{dc}
\frac {\partial}{\partial \psii Bd}\quad ,}{ }
where
\eqn{D^{AA'}{}_{ab}=\frac i{\sigma h^{\frac 1 2}}h_{ab}n^{AA'}+
\frac 1{\sigma} \epsilon_{abc}\Omega^{AA'c}}{ }
and $\sigma =\int \d^3x E$.
We have chosen the operator ordering in $p_{AA'}{}^a$ so that it becomes
 hermitian with respect to the formal inner product [\DE].

Note that we can also take the wave function as an eigenstate of
 $\psibari Aa$ and $\Omega^{AA'}{}_a$ : $\tilde{\Psi}=\tilde{\Psi}
(\Omega ^{AA'}{}_a, \psibari Aa)$, which relates to $\Psi(\Omega ^{AA'}{}_a,
 \psii Aa)$ by the fermionic Fourier transform
\eqn{\tilde{\Psi}(\Omega ^{AA'}{}_a, \psibari Aa)=D^{-1}(\Omega)\int
 \Psi (\Omega ^{AA'}{}_a, \psii Aa)\exp
\paren{-\frac i \hbar C_{AA'}{}^{ab}\psii Aa \psibari Ab }
\prod_{E,e}\d\psii Ee,}{fo1}
where
\eqn{C_{AA'}{}^{ab} = -\sigma \epsilon^{abc}\Omega_{AA'c}\quad ,}{ }
\eqn{D(\Omega)=\det\paren{-\frac i \hbar C_{AA'}{}^{ab}}.}{ }
With this relation, we can obtain $\tilde{\Psi}$ from $\Psi$, and vice versa.

Physically allowed wave functions are obtained by solving quantum versions
 of constraint equations:
\eqn{J_{AB}\Psi =0 \enskip , \qquad \bar J_{A'B'}\Psi =0 \enskip,}{j1}
\eqn{S_A \Psi =0 \enskip, \qquad \bar S_{A'}\Psi =0\enskip ,}{s2}
\eqn{H_{\bot}\Psi =0 \enskip, \qquad H_a\Psi =0 \enskip .}{h3}
Because of the relation \{$S_A$, $\bar S_{A'}$\} $\propto$
$ H_{\bot}n_{AA'}+H_a\Omega_{AA'}{}^a$ up to terms proportional to $J$ and
 $\bar J$, eqs. \reff{h3} hold automatically if the conditions \reff{j1}
 and \reff{s2} are satisfied.
The constraints \reff{j1} describe the invariance of $\Psi$ under Lorentz
 transformations.
We therefore only have to solve $S_A\Psi =0$ and $\bar S_{A'}\Psi =0$ under
 the assumption that $\Psi$ is Lorentz invariant.
Let us write the explicit forms of $S_A$ and $\bar S_{A'}$:
\eqn{\bar S_{A'}=\sigma m^{ab}\Omega_{AA'a}\psii Ab+
\rcp2 i\kappa^2\paren{s p_{AA'}{}^a\psii Aa +(1-s)\psii Aa p_{AA'}{}^a},}
{asano-24}
\eqn{S_A=\sigma m^{ab}\Omega_{AA'a}\psibari {A} b -\rcp2
i\kappa^2\paren{(1-s) p_{AA'}{}^a\psibari Aa +s \psibari Aa p_{AA'}{}^a},}
{asano-25}
where $\kappa^2=8\pi$ and $s$ parametrizes the ambiguity of the operator
 ordering, which comes from noncommutativity of $\psii Aa,\, \psibari Aa $
 and $p_{AA'}{}^a$.
Note that the orderings in refs. [\DE] and [\DHO] can be obtained by taking
 $s$ to be $0$ and $\frac 1 2$, respectively.
Now we write, without specifying the value of $s$, the constraint equations
 \reff{s2} in an explicit form
\eqn{\paren{\gamma m^{ab}\Omega_{AA'a}\psii Ab+\psii Aa \frac{\partial}
{\partial \Omega^{AA'}{}_a}+s\psii Aa\Omega _{AA'}{}^a}\Psi =0,}{s6}
\eqn{\paren{-\gamma m^{ab} \Omega_{AA'a}D^{BA'}{}_{cb}
\frac{\partial}{\partial \psii Bc}+\frac{\partial}{\partial\Omega^{AA'}
{}_a}\paren{D^{BA'}{}_{ba}\frac{\partial}{\partial \psii Bb}}-s\Omega_{AA'}
{}^aD^{BA'}{}_{ba}\frac{\partial}{\partial \psii Bb}}\Psi=0,}{s7}
where $\gamma \equiv {2\sigma}/{\hbar \kappa^2}$.
Moreover, using the Fourier transform \reff{fo1}, eqs. \reff{s6} and
 \reff{s7} can be recast into the conditions for
$\tilde{\Psi}(\Omega ^{AA'}{}_a, \psibari Aa)$ :
\eqn{\paren{-\gamma m^{ab} \Omega_{AA'a}D^{AB'}{}_{bc}
\frac{\partial}{\partial \psibari Bc}-\frac{\partial}{\partial\Omega^{AA'}
{}_a}\paren{D^{AB'}{}_{ab}\frac{\partial}{\partial \psibari Bb}}+
(2-s)\Omega_{AA'}{}^aD^{AB'}{}_{ab}\frac{\partial}{\partial \psibari Bb}}
\tilde{\Psi}=0,}{til1}
\eqn{\paren{\gamma m^{ab}\Omega_{AA'a}\psibari Ab-\psibari Aa
\frac{\partial}{\partial \Omega^{AA'}{}_a}+s\psibari Aa\Omega _{AA'}{}^a}
\tilde{\Psi} =0.}{til2}

\section{Solving quantum constraints}

Because of the Lorentz invariant property,
 $\Psi$ can only contain an even number of $\psii Aa$.
Thus, we can decompose $\Psi$ into $\psi^0$, $\psi^2$, $\psi^4$,
$\psi^6$ parts, which we write as $\Psi_0$, $\Psi_2$, $\Psi_4$,
$\Psi_6$, respectively, so that we have $\Psi = \Psi_0+ \Psi_2 +\Psi_4+\Psi_6$.
These terms are related to the $\bar\psi^6$, $\bar\psi^4$, $\bar\psi^2$,
$\bar\psi^0$ parts of $\tilde\Psi$ by the Fourier transform, respectively.
We write $\tilde\Psi = \tilde\Psi_6+ \tilde\Psi_4 +\tilde\Psi_2+
\tilde\Psi_0$, where the subscripts represent the number of
 $\psibari Aa$ contained in each term.

We shall give these terms separately by solving the constraint equations
 along lines similar to ref. [\DHO].

For $\Psi _0$, eq. \reff{s7} is trivial because $\Psi_0$ has no fermionic
 variables.
Bearing in mind that Lorentz invariance imposes a condition
$\Psi_0 = \Psi_0 (h_{ab})$, eq. \reff{s6} reduces to
\eqn{\paren {\gamma m^{ab}-2\frac \partial {\partial h_{ab}} +
sh^{ab}}\Psi_0 =0,}{ }
so that we obtain
\eqn{\Psi _0 =const. h^{\frac s{2}}\exp \paren{\frac 1 2
 \gamma m^{ab}h_{ab}},}{asano-32}
where $h \equiv \det (h_{ab})$.

Next, let us look at $\Psi_2$.
Solving eqs.\reff{s6} and \reff{s7} under the assumption that
 $\Psi_2$ is Lorentz invariant, we get
\eqn{\Psi_2 =0 }{asano-34}
as solutions except for $s=0,-4/3$ and $m^{ab}=0$.
In these exceptional cases, there are the formal solutions
\eqna{\Psi_2 \wa const. \psi''{}^2 \quad (s=0),\\
      \Psi_2 \wa const. (\psi'{}^2+2\psi''{}^2) \quad (s=-4/3) , }
where $\psi'{}^2\equiv \psi_A{}^a\psi^A{}_a$ and
$\psi''{}^2\equiv ih^{-1/2}\epsilon^{abc}\Omega_{AA'c}n_B{}^{A'}
\psi^A{}_a\psi^B{}_b$.
However, since they cease to be solutions when considering a small
 perturbation, we exclude them from our consideration.

For $\Psi_4$, or equivalently, $\tilde\Psi_2$, using a similar method to
 the case of $\Psi_2$, eqs. \reff{til1} and \reff{til2} can be solved
 to yield $\tilde\Psi_2=0$, i.e.,
\eqn{\Psi _4=0.}{asano-36}
Although there are also exceptional formal solutions when $s=0,10/3$ and
 $m^{ab}=0$, we discard them for the same reason as above.

For $\tilde\Psi_0$, the solution is easily found to be
\eqn{\tilde\Psi_0=const.h^{-\frac {s}{2}}\exp \paren{-\frac 1 2
 \gamma m^{ab}h_{ab}}}{ }
as in the case of $\Psi_0$. In other words, we have
\eqn{\Psi _6 =const. h^{\frac {2-s}{2}}\exp \paren{-\frac 1 2
\gamma m^{ab}h_{ab}}(\psii Aa \psi_{Aa})^3,}{asano-38}
where $\psii Aa\psi_{Aa}=\psii Aa\psi_{Ab}\delta ^{ab}$.

It is rather remarkable that all the solutions \reff{danii-1} for Bianchi
 class A models are found in a unified way.
We emphasize that these solutions are actually exact.

Recently, D'Eath claimed that the full $N=1$ supergravity has no quantum
 corrections [\DEpre].
The reduced $N=1$ supergravity presented above also has no quantum corrections
 and looks like a free theory.
If one wants to see the effect of interactions, one probably has to explore
 $N=1$ supergravity coupled to supermatter.

\section{An example: Bianchi type IX case}

One specific case would be of particular interest, i.e. Bianchi type
 IX representing homogeneous $S^3$ universe.

Giving the invariant basis for it as
\eqn{E^1 = \sin\beta\cos\gamma\d\alpha-\sin\gamma\d\beta,\;
      E^2 = \sin\beta\sin\gamma\d\alpha+\cos\gamma\d\beta,\;
      E^3 = \cos\beta\d\alpha+\d\gamma,}{2-1}
where $E^a\equiv E^a{}_i\d x^i,\, x^1=\alpha,\, x^2=\beta\,{\rm and}\,
x^3=\gamma$,
the structure constants are given by $m^{ab}=\diag{1,1,1}$.
And taking the range of variables as
\eqn{(0\leq\alpha\leq4\pi,\quad 0\leq\beta\leq\pi,
	\quad 0\leq\gamma\leq2\pi),}{2-2}
we can easily find $\sigma=16\pi^2$, so that we have
\eqn{I\equiv\rcp2\gamma m^{ab}\dh ab={2\pi\over\hbar}{\rm Tr}(\dh ab).}{2-5}

For the semi-classical forms of the quantum states,
it seems natural to ask if the states
$\Psi_0$ and $\Psi_6$ are the Hartle-Hawking state
[\HH] and the wormhole state [\HP], respectively.
To answer this question, we note that vacuum Bianchi type IX metrics are
diagonalizable [\RS].
In such cases it is fortunately known [\BGPP,\GP] that the asymptotically
Euclidean four-metrics which correspond to
the classical flow of $I$ are found to be
\eqn{\d s^2=F^{-1/2}\d\rho^2+F^{1/2}{\rho^2\over4}
\bra{\paren{1-\frac{a_1^4}{\rho^4}}\inv\! (E^1)^2+
     \paren{1-\frac{a_2^4}{\rho^4}}\inv\! (E^2)^2+
     \paren{1-\frac{a_3^4}{\rho^4}}\inv\! (E^3)^2},
}{r1}
where
\eqn{F=\paren{1-\frac{a_1^4}{\rho^4}}\paren{1-\frac{a_2^4}{\rho^4}}
\paren{1-\frac{a_3^4}{\rho^4}}
}{r2}
and $a_1,a_2,a_3$ are constants.
At large distances, the metrics \reff{r1} are well-behaved and $\Psi_6$ dumps
rapidly.
Hence $\Psi_6$ is certainly the ground quantum wormhole state.
On the other hand, though $\Psi_0$ itself is regular at small three-geometries,
the classical flow of $I$ is singular when $\rho\goes 0$.
This means the quantum state $\Psi_0$ is {\it not}
the Hartle-Hawking state [\DErev].

The `filled' state $\Psi_6$ corresponds to the wormhole state in
our description above and this seems to be in contradiction with the literature
[\GrPRL,\GrPL,\DErev], where the bosonic state $\Psi_0$ corresponds
to the wormhole state.
However, another possibility exists because of the
indefiniteness of the
 sign of $\sigma$, which is defined as $\int \d^3\! xE$, not $\int \d^3\!
 x\abs E$.
In fact, by the following antipodal transformation
\eqn{\beta\goes\beta+\pi,}{2-10}
we have $\sigma\goes-\sigma, \ m^{ab}\goes m^{ab}$, so that $I\goes -I$.
Correspondingly this time the bosonic state $\Psi_0$ becomes the wormhole
state.
We conclude that there are two distinct solutions of the constraints in
 the case of Bianchi type IX, which are mutually related by the antipodal
transformation.

\bigskip

\section*{Acknowledgments}

We would like to thank Prof. A. Hosoya for helpful discussions
and Dr. P.D. D'Eath for correspondence.

\section*{References}

\noindent [\HK] See, e.g., K. Hagiwara, S. Komamiya, in High energy
 electron-positron physics ed. by A. Ali and P. S\"{o}ding
 (World Scientific, Singapore, 1988).

\noindent [\RS] M.P. Ryan and L.C. Shepley, Homogeneous relativistic
 cosmologies (Princeton U.P., Princeton, NJ, 1975).

\noindent [\GrPRL] R. Graham, Phys. Rev. Lett. 67 (1991) 1381.

\noindent [\GrPL] R. Graham, Phys. Lett. B 277 (1992) 393.

\noindent [\DHO] P.D. D'Eath, S.W. Hawking and O. Obreg\'{o}n, Phys. Lett.
B 300 (1993) 44.

\noindent [\DE] P.D. D'Eath, Phys. Rev. D 29 (1984) 2199.

\noindent [\DEpre] P.D. D'Eath, preprint, `Physical states in $N=1$
supergravity',
(1993).

\noindent [\HH] J.B. Hartle and S.W. Hawking, Phys. Rev. D 28 (1983) 2960.

\noindent [\HP] S.W. Hawking and D.N. Page, Phys. Rev. D 42 (1990) 2655.

\noindent [\BGPP] V.A. Belinskii, G.W. Gibbons, D.N. Page and C.N. Pope,
Phys. Lett. B 76 (1978) 433.

\noindent [\GP] G.W. Gibbons and C.N. Pope, Comm. Math. Phys. 66 (1979) 267.

\noindent [\DErev] P.D. D'Eath, preprint,
`Quantization of the Bianchi IX model in supergravity', (1993), to appear in
Phys. Rev. D.

\noindent [\OSB] O. Obreg\'{o}n, J. Socorro and J. Ben\'{i}tez, Phys. Rev. D
47 (1993) 4471.

\noindent [\KTH] T. Koike, M. Tanimoto and A. Hosoya, preprint,
`Compact Homogeneous Universes', TIT/HEP-208/COSMO-26 Dec. (1992).

\end{document}